\documentstyle[12pt]{article}
\expandafter\ifx\csname amssym.def\endcsname\relax \else\endinput\fi
\expandafter\edef\csname amssym.def\endcsname{%
       \catcode`\noexpand\@=\the\catcode`\@\space}
\catcode`\@=11
\def\undefine#1{\let#1\undefined}
\def\newsymbol#1#2#3#4#5{\let\next@\relax
 \ifnum#2=\@ne\let\next@\msafam@\else
 \ifnum#2=\tw@\let\next@\msbfam@\fi\fi
 \mathchardef#1="#3\next@#4#5}
\def\mathhexbox@#1#2#3{\relax
 \ifmmode\mathpalette{}{\m@th\mathchar"#1#2#3}%
 \else\leavevmode\hbox{$\m@th\mathchar"#1#2#3$}\fi}
\def\hexnumber@#1{\ifcase#1 0\or 1\or 2\or 3\or 4\or 5\or 6\or 7\or 8\or
 9\or A\or B\or C\or D\or E\or F\fi}

\font\tenmsa=msam10 scaled\magstep1
\font\sevenmsa=msam7 scaled\magstep1
\font\fivemsa=msam5 scaled\magstep1
\newfam\msafam
\textfont\msafam=\tenmsa
\scriptfont\msafam=\sevenmsa
\scriptscriptfont\msafam=\fivemsa
\edef\msafam@{\hexnumber@\msafam}
\mathchardef\dabar@"0\msafam@39
\def\dashrightarrow{\mathrel{\dabar@\dabar@\mathchar"0\msafam@4B}}
\def\dashleftarrow{\mathrel{\mathchar"0\msafam@4C\dabar@\dabar@}}

\def\ulcorner{\delimiter"4\msafam@70\msafam@70 }
\def\urcorner{\delimiter"5\msafam@71\msafam@71 }
\def\llcorner{\delimiter"4\msafam@78\msafam@78 }
\def\lrcorner{\delimiter"5\msafam@79\msafam@79 }
\def\yen{{\mathhexbox@\msafam@55 }}
\def\checkmark{{\mathhexbox@\msafam@58 }}
\def\circledR{{\mathhexbox@\msafam@72 }}
\def\maltese{{\mathhexbox@\msafam@7A }}

\font\tenmsb=msbm10 scaled\magstep1
\font\sevenmsb=msbm7 scaled\magstep1
\font\fivemsb=msbm5 scaled\magstep1 
\newfam\msbfam
\textfont\msbfam=\tenmsb
\scriptfont\msbfam=\sevenmsb
\scriptscriptfont\msbfam=\fivemsb
\edef\msbfam@{\hexnumber@\msbfam}
\def\Bbb#1{{\fam\msbfam\relax#1}}
\def\widehat#1{\setbox\z@\hbox{$\m@th#1$}%
 \ifdim\wd\z@>\tw@ em\mathaccent"0\msbfam@5B{#1}%
 \else\mathaccent"0362{#1}\fi}
\def\widetilde#1{\setbox\z@\hbox{$\m@th#1$}%
 \ifdim\wd\z@>\tw@ em\mathaccent"0\msbfam@5D{#1}%
 \else\mathaccent"0365{#1}\fi}
\font\teneufm=eufm10 scaled\magstep1
\font\seveneufm=eufm7 scaled\magstep1
\font\fiveeufm=eufm5 scaled\magstep1
\newfam\eufmfam
\textfont\eufmfam=\teneufm
\scriptfont\eufmfam=\seveneufm
\scriptscriptfont\eufmfam=\fiveeufm

\csname amssym.def\endcsname

\input{amssym.tex}
\begin{document}
\input {epsf}

\def\ov{\overline}
\def\onlyif{\rightarrow}
\def\and{\th\th\&\th\th}
\def\iff{\leftrightarrow}
\def\dv{\vdash}
\def\ddv{\dashv\vdash}
\def\openone{\leavevmode\hbox{\small1\kern-3.8pt\normalsize1}}
\def\G{G\"odel}
\def\Schr{Schr\"odinger}
\def\na{na\"\i ve}
\def\maczy{M\c{a}czy\'{n}ski}
\def\sing{\Psi_S}
\def\state{\Psi}
\def\spinvec{{\bf\sigma}}
\def\spina{\sigma_a}
\def\spinb{\sigma_b}
\def\spinc{\sigma_c}
\def\spinx{\sigma_x}
\def\spiny{\sigma_y}
\def\spinz{\sigma_z}
\def\unita{{\bf\hat a}}
\def\unitb{{\bf\hat b}}
\def\unitc{{\bf\hat c}}
\def\unitz{{\bf\hat z}}
\def\identity{{\bf I}}
\def\a{\alpha}
\def\b{\beta}
\def\g{\gamma}
\def\r{\rho}
\def\minus{\,-\,}
\def\eks{\bf x}
\def\kay{\bf k}
\def\lorentz{
  \sqrt{1 - \beta^2} }

\def\ket#1{|\,#1\,\rangle}
\def\bra#1{\langle\, #1\,|}
\def\braket#1#2{\langle\, #1\,|\,#2\,\rangle}
\def\proj#1#2{\ket{#1}\bra{#2}}
\def\expect#1{\langle\, #1\, \rangle}
\def\trialexpect#1{\expect#1_{\rm trial}}
\def\ensemblexpect#1{\expect#1_{\rm ensemble}}
\def\kpsi{\ket{\psi}}
\def\kphi{\ket{\phi}}
\def\bpsi{\bra{\psi}}
\def\bphi{\bra{\phi}}

\def\ditto{\rule[0.5ex]{2cm}{.4pt}\enspace}
\def\th{\thinspace}
\def\ni{\noindent}
\def\thirty{\hbox to \hsize{\hfill\rule[5pt]{2.5cm}{0.5pt}\hfill}}

\def\half{\frac{1}{2}}
\def\third{\frac{1}{3}}
\def\squarthtwo{\frac{1}{\sqrt 2}}
\def\squarth#1{\frac{1}{\sqrt{#1}}}
\def\cubth#1{\frac{1}{^3\sqrt{#1}}}

\def\set#1{\{ #1\}}
\def\setbuilder#1#2{\{ #1:\; #2\}}
\def\Prob#1{{\rm Prob}(#1)}
\def\pair#1#2{\langle #1,#2\rangle}
\def\Id{\bf 1}

\def\dee#1#2{\frac{\partial #1}{\partial #2}}
\def\deetwo#1#2{\frac{\partial\,^2 #1}{\partial #2^2}}
\def\deethree#1#2{\frac{\partial\,^3 #1}{\partial #2^3}}

\def\openone{\leavevmode\hbox{\small1\kern-3.8pt\normalsize1}}
\normalsize
\title{Nonlinear quantum state transformation of
spin-$1/2$}
\author{
H. Bechmann-Pasquinucci, B. Huttner, N. Gisin \\
\small
{\it Group of Applied Physics, University of Geneva, CH-1211, Geneva 4,
Switzerland}}
\maketitle
\abstract{
A non-linear quantum state transformation is presented. The
transformation, which operates on pairs of spin-1/2, can be
used to distinguish optimally between two non-orthogonal states. Similar
transformations applied locally on each component of an entangled pair of
spin-1/2 can be used
to transform a mixed nonlocal state into a quasi-pure maximally entangled
singlet state. In both cases the transformation makes use of the basic
building block of the quantum computer, namely the quantum-XOR gate. 
}

\normalsize
\section{Introduction}
Consider the following transformation of a spin-$1/2$ density matrix:
\begin{eqnarray}
{\rho}^{in} = \left( \begin{array}{cc} {\rho}_{11}& {\rho}_{12} \\       
{\rho}_{21} &{\rho}_{22} \end{array} \right) ~~~ \longrightarrow ~~~
{\rho}^{out} = \left( \begin{array}{cc} {({\rho}_{11})}^{2}&
{({\rho}_{12})}^{2}
\\
{({\rho}_{21})}^{2} &{({\rho}_{22})}^{2} \end{array} \right)
\label{eq:sqrho}
\end{eqnarray}
The question is if this transformation corresponds to a physically
feasible quantum state transformation. At first one may be tempted to
say:"of course not! This transformation is nonlinear and does not preserve
the trace". But it turns out that this transformation is indeed possible
to realize in the lab, it is only a question
of technological difficulties and therefore of time. This transformation,
which involves the
fundamental building component of a quantum computer, can be used to
differentiate optimally between non-orthogonal spin-$1/2$ states. A
similar
transformation applied to entangled pairs of spin-$1/2$ can be used to
transform a
mixed nonlocal state into a quasi-pure maximally entangled singlet
state.

Before constructing such a state transformer (using unitary and other well
accepted transformations), it should be stressed that the output
${\r}^{out}$ depends only on ${\r}^{in}$, not on any decomposition of
${\r}^{in}$ into pure states. Therefore, this kind of
non-linear transformation does not lead to the 'arbitrary fast signalling'
problem \cite{gisin1,gisin2}. 
That the trace is not preserved does not give rise to any problems either,
since it merely reflects that some spins are lost during the
transformation. If desirable, one could simply renormalize ${\r}^{out}$.

In the next section it is shown how the transformation eq.
(\ref{eq:sqrho}) comes about.
In  Sec.~\ref{sec:trid}, it will be shown how the
non-linear transformation can be used to distinguish between two
non-orthogonal states, provided two copies of the state are available. In
Sec.~\ref{sec:lige} the Loss Induced Generalized (LIGe) quantum
measurement is presented \cite{huttner}. It is a special construction of
Positive
Operator Value Measure (POVM), and it is shown how this specific
measurement can be applied in two different ways, when two copies of the
state are provided. It turns out that the 3 different ways to
distinguish between two non-orthogonal states which will be described
here all leads to the same probability of successfully determining
the input state. In Sec.~\ref{sec:opti} it is proven that the probability
of successfully determining the state is indeed also the optimal solution
to the state identification problem.

A generalized non-linear transformation, operating on entangled
pairs of spin-1/2, can be used to transform a  mixed nonlocal state
into a quasi-pure maximally entangled singlet state. This will be
described in Sec.~\ref{sec:puri}.

\section{How it works}
\label{sec:how}
To realize the transformation, one needs at least two identical
independent copies of
the same state ${\rho}^{in}$, and it will be assumed that this is
possible.

{\it The first step} consists of considering the spins pairwise,
\begin{eqnarray}
{\rho}^{in} ~~~ \longrightarrow ~~~{\rho}^{in} \otimes {\rho}^{in}
\end{eqnarray}
This is quite easy to do, since nothing needs to be done
physically. It is however this crucial step which makes the final
transformation nonlinear. 

{\it The second step} consists of a 'controlled not
gate' interaction between each spin in a pair. This is physically the
hard part of the whole transformation. A controlled not gate (or  
quantum-XOR) flips the second spin (target spin) if and only if the
first
(source spin) is 'spin-up'. It is a unitary transformation, ${\Bbb
{U}}_{\rm XOR}$,
acting on pairs of spin-$1/2$:
\begin{eqnarray}
\begin{array}{lll}
\ket{++} & \longrightarrow & \ket{+-}\\ 
\ket{+-} & \longrightarrow & \ket{++}\\
\ket{-+} & \longrightarrow & \ket{-+}\\
\ket{--} & \longrightarrow & \ket{--}\\ \end{array}
\end{eqnarray}
or when written in matrix notation:
\begin{eqnarray}
{{\Bbb {U}}_{\rm XOR}}=
\left( \begin{array}{cc} \spinx & 0\\ 0 & {\openone} \end{array} \right)
\label{eq:xor}
\end{eqnarray}
where $\spinx = \left( \begin{array}{cc} 0 & 1\\ 1 & 0 \end{array}\right)$
is the first Pauli matrix and ${\openone}$ is the $2\times 2$ identity 
matrix. Note that the XOR, is the basic building
block for a quantum processor \cite{barenco}.

{\it The third step} is again easy: measure the spin component of the
second particle along the
$z$-direction and keep the pair only if
the result is 'down'. Hence there is a probability of failure.

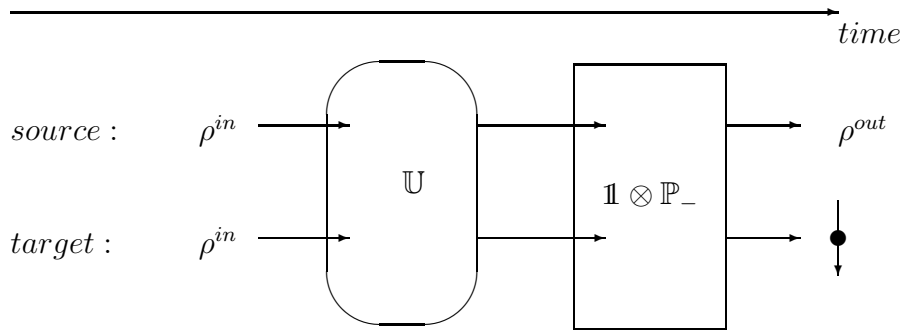
\begin{figure}[t]
\begin{center}
\begin{picture}(110,50)(-25,0)
\put(-25,47){\vector(1,0){110}}
\put(85,43){$time$}
\put(-25,15){$target:$} 
\put(-25,30){$source:$}
\put(0,15){${\rho}^{in}$}
\put(0,30){${\rho}^{in}$}
\put(8,17){\vector(1,0){12}}
\put(8,32){\vector(1,0){12}}
\put(27,23){\oval(20,35){$\Bbb U$}}
\put(37,17){\vector(1,0){17}}
\put(37,32){\vector(1,0){17}}
\put(50,5){\framebox(20,35){${\openone \otimes {\Bbb P}_{-}}$}}
\put(70,17){\vector(1,0){10}}
\put(70,32){\vector(1,0){10}}
\put(85,30){${\rho}^{out}$}
\put(85,22){\vector(0,-1){10}}
\put(85,17){\circle*{2}}
\end{picture}
\end{center}
\caption{Here the process is schematically outlined.
The two copies of the
same spin state first undergo an unitary interaction, and afterwards a
filtering process. The source spin ${\rho}^{in}$ will afterwards be in a
new state ${\rho}^{out}$,
whereas the state of the target spin always is reduced to a spin 'down'
state.}
\label{fig:trans}
\end{figure}

Altogether (see Fig.\ref{fig:trans}) this procedure transforms the
state of the source spin from ${\rho}^{in}_{source}\rightarrow
{\rho}^{out}_{source}$, whereas
the target spin is always left in the spin-down state
${\rho}^{in}_{target}\rightarrow
{\rho}^{out}_{target}=\ket{-}\bra{-}={\Bbb P}_{-}$. The whole
transformation can be written as 
\begin{eqnarray}
 \left({\openone}\otimes
{\Bbb P}_{-} \left({\Bbb U}_{\rm XOR} \left({\r}^{in}\otimes{\r}^{in} 
\right){\Bbb U}_{\rm XOR}^{\dag} \right){\openone}\otimes
{\Bbb P}_{-}\right)={\r}^{out}\otimes
{\Bbb P}_{-}
\label{eq:trans}
\end{eqnarray}
Where ${\r}^{out}$ is the density matrix in eq. (\ref{eq:sqrho}), where
each matrix element has been squared by itself.
 
Other similar transformations can be build by using other components,
which means substituting the control-not gate with other unitary
interactions, see Fig. 2.

\section{Using the transformation for state identification}
\label{sec:trid}
The non-linear transformation which has just been described can be used to
transform non-orthogonal states into orthogonal states, provided two
copies of the state are available. By nature, non-orthogonal quantum
states
can not be distinguished with certainty, and even if this transformation
can turn non-orthogonal states into orthogonal states, this does not in
any
way conflict with quantum mechanics, since there is a certain
probability that the transformation fails.

A density matrix describing a spin-1/2 system can be represented on the
Poincare
sphere in terms of a polarization vector ${{P}}=(x,y,z)$ (also
known as Bloch vector) and the
Pauli-matrices ${{\sigma}}=({\sigma}_{x},{\sigma}_{y},{\sigma}_{z})$ 
in the following way:
\begin{eqnarray}
\r = \frac{1}{2}({\openone}+ {P}\cdot{\sigma})
\end{eqnarray}

For two spin-1/2 states to be orthogonal their corresponding polarization
vectors must point in opposite directions on the sphere, in other words   
two states $\ket{a}$ and $\ket{b}$ are orthogonal if their polarization
vectors satisfy ${P}_{a}=-{P}_{b}$.

Consider the two pure spin-1/2 states $\ket{{\psi}_{1}^{in}}$ and
$\ket{{\psi}_{2}^{in}}$ with the polarization vectors
${P}^{in}_{1}$ and ${P}^{in}_{2}$ expressed in the usual
spherical coordinates
\begin{eqnarray}
\begin{array}{ll}
{x}_{i}^{in} = \sin{\theta}_{i} \cos{\phi}_{i} \\
{y}_{i}^{in} = \sin{\theta}_{i} \sin{\phi}_{i} \\
{z}_{i}^{in} = \cos{\theta}_{i}\\
\end{array}
\end{eqnarray}
where $i=1,2$. When transforming these two states, which means taking two
copies of the
same state,
performing the control-not gate interaction between them and the
filtering measurement, the new state have the
following polarization vectors ${P}^{out}_{i}$:\footnote{Notice that the
outgoing polarization vectors 
${P}^{out}_{i}$ are not normalized}
\begin{eqnarray}
\begin{array}{ll}
{x}_{i}^{out} =\frac{1}{2} {\sin}^{2}{\theta}_{i} \cos{2\phi}_{i} \\
{y}_{i}^{out} =\frac{1}{2} {\sin}^{2}{\theta}_{i} \sin{2\phi}_{i} \\
{z}_{i}^{out} = \cos{\theta}_{i} \\
\end{array}
\end{eqnarray}
For the corresponding spin states $\ket{{\psi}_{1}^{out}}$ and
$\ket{{\psi}_{2}^{out}}$ to be
orthogonal the 
requirement is that their corresponding polarization vectors must be
opposite, i.e. ${P}^{out}_{1}=-{P}^{out}_{2}$. This can be
fulfilled by imposing the following relations between the angles;
\begin{eqnarray}
&&\cos{\theta}_{1}=-\cos{\theta}_{2} \Longrightarrow
{\theta}_{2}={\theta}_{1}+\pi ~~~~~{\theta}\equiv
{\theta}_{1}={\theta}_{2}-\pi \nonumber \\
&&\cos{2\phi}_{1}=-\cos{2\phi}_{2} ~~~{\rm and} 
~~~\sin{2\phi}_{1}=-\sin{2\phi}_{2} ~~~~~ {\phi}\equiv
{\phi}_{1}={\phi}_{2}-{\frac{\pi}{2}}\nonumber 
\end{eqnarray}
Imposing these constrains imply that the initial states
$\ket{{\psi}_{1}^{in}}$ and
$\ket{{\psi}_{2}^{in}}$ had the following
density matrices,
\begin{eqnarray}
 & & {\r}_{1}^{in}={\frac{1}{2}}
\left( \begin{array}{cc} 1+{\cos}{\theta} & {\sin}{\theta}{e}^{-i\phi}\\ 
{\sin}{\theta}{e}^{i\phi}& 1-{\cos}{\theta} \end{array} \right)
=\ket{{\psi}_{1}^{in}}\bra{{\psi}_{1}^{in}} \\
 & & {\r}_{2}^{in}={\frac{1}{2}}
\left( \begin{array}{cc} 1-{\cos}{\theta} & {i \sin}{\theta}{e}^{-i\phi}\\
{-i \sin}{\theta}{e}^{i\phi}& 1+{\cos}{\theta} \end{array} \right)
=\ket{{\psi}_{2}^{in}}\bra{{\psi}_{2}^{in}}\nonumber
\end{eqnarray}
Whereas the transformed states  $\ket{{\psi}_{1}^{out}}$ and
$\ket{{\psi}_{2}^{out}}$ have as density matrices  the input matrices
with each matrix element squared by itself, as seen in eq.
(\ref{eq:sqrho}).

The two initial states are not orthogonal, i.e.
${\r}_{1}^{in}{\r}_{2}^{in}\neq
0$ which means that they can not be distinguished with certainty. The
overlap between the states, which can be obtained from
${\r}_{1}^{in}{\r}_{2}^{in}{\r}_{1}^{in}=
{|\braket{{\psi}_{1}^{in}}{{\psi}_{2}^{in}}|}^{2}{\r}_{1}^{in}$,
is found to be  
\begin{eqnarray}
|\braket{{\psi}_{1}^{in}}{{\psi}_{2}^{in}}|=
\frac{{\sin}{\theta}}{\sqrt{2}}
\end{eqnarray} 
The two outgoing states are orthogonal, i.e. 
${\r}_{1}^{out} {\r}_{2}^{out} =0$. This means that they can
now be
identified with certainty by performing a standard von Neumann 
measurement.
If the
state $\ket{{\psi}_{1}^{out}}$ is obtained the initial states was
$\ket{{\psi}_{1}^{in}}$. Similarilly, if the state
$\ket{{\psi}_{2}^{out}}$ is found
the initial state was $\ket{{\psi}_{2}^{in}}$. 
In other words the
transformation makes it possible to distinguish {\it with certainty}
between states which were originally not distinguishable. 
However, the transformation is
not always successful. There is, in fact,  only a certain probability that
it will succeed, which is given by the trace of the ${\r}_{i}^{out}$,
where $i$ is either $1$ or $2$. 
It is important to realize that when the transformation is
successful, the initial state is completely identified. Whereas if the
transformation fails, no knowledge about the initial state can be
obtained. 

The total probability of successfully identifying the state is 
\begin{eqnarray}
{\rm Pr}_{\rm T}(success) = {\frac{1}{2}}\left( {\rm
Tr}({\r}_{1}^{out})
+ {~\rm Tr}\left( {\r}_{2}^{out}\right) \right) = 1
-{\frac{{\sin}^{2}{\theta}}{2}}
\end{eqnarray}
Notice that this is equal to $1
- {|\braket{{\psi}_{1}^{in}}{{\psi}_{2}^{in}}|}^{2}$.

\section{Another way to identify non-orthogonal states}
\label{sec:lige}
There are other ways of distinguishing between two non-orthogonal states
without making any errors. A well-known way is to use 
Positive Operator Value
Measure (POVM), which are generalized quantum measurements. 
The Loss Induced Generalized
(LIGe) quantum measurement is a special POVM \cite{ivan,peres1}, which
has even been
performed
experimentally \cite{huttner}. In the case of LIGe the idea is
that the two non-orthogonal spin-$1/2$ lie in a plane spanned by two
orthogonal states $\ket{{\phi}_{1}}$ and $\ket{{\phi}_{2}}$,
i.e.
\begin{eqnarray}
 &&\ket{{\psi}^{in}_{1}}={\cos}{\frac{\alpha}{2}}\ket{{\phi}_{1}}+ 
{\sin}{\frac{\alpha}{2}}\ket{{\phi}_{2}}\\
 &&\ket{{\psi}^{in}_{2}}={\cos}{\frac{\alpha}{2}}\ket{{\phi}_{1}}-
{\sin}{\frac{\alpha}{2}}\ket{{\phi}_{2}}\nonumber \end{eqnarray}
with the overlap $|\braket{{\psi}_{1}^{in}}{{\psi}_{2}^{in}}|=
{\cos}{\alpha}$. The procedure is to add one dimension
$\ket{{\phi}_{0}}$ orthogonal to $\ket{{\phi}_{1}}$ and
$\ket{{\phi}_{2}}$, and then perform a rotation around $\ket{u}\equiv 
\ket{{\phi}_{1}} - \ket{{\phi}_{2}}$, with an angle $\cos{\theta}=\tan{
\frac{\alpha}{2}}$. After  the rotation the two states can be written as
\begin{eqnarray}
 & &\ket{{\psi}^{out}_{1}}=
\sqrt{2}\sin{\frac{\alpha}{2}}\ket{{\phi}_{1}}+ 
{\sqrt{\cos\alpha}}\ket{{\phi}_{0}}\\
 & &\ket{{\psi}^{out}_{2}}=
\sqrt{2}\sin{\frac{\alpha}{2}}\ket{{\phi}_{2}}+
{\sqrt{\cos\alpha}}\ket{{\phi}_{0}}\nonumber
\end{eqnarray}
Since the three states $\ket{{\phi}_{0}}$, $\ket{{\phi}_{1}}$ and
$\ket{{\phi}_{2}}$ are orthogonal they can be separated deterministically
with a standard measurement. 

If the initial state was $\ket{{\psi}^{in}_{1}}$ the result of a
measurement
will either be $\ket{{\phi}_{1}}$ with the probability
$2{\sin}^{2}{\frac{\alpha}{2}}$ or $\ket{{\phi}_{0}}$ with probability 
$\cos \alpha$. Where as if the initial state was $\ket{{\psi}^{in}_{2}}$
the
result of a measurement will either be $\ket{{\phi}_{2}}$ with the
probability $2{\sin}^{2}{\frac{\alpha}{2}}$ or $\ket{{\phi}_{0}}$ with
probability $\cos \alpha$.

Therefore when the state obtained is either $\ket{{\phi}_{1}}$ or
$\ket{{\phi}_{2}}$, we can conclude that the initial state  was
$\ket{{\psi}^{in}_{1}}$ and $\ket{{\psi}^{in}_{2}}$ respectively. Whereas
if the
obtained state is $\ket{{\phi}_{0}}$ the initial state could have been
either of the two and in order not to introduce any errors these results
are discarded. The
probability of identifying the state is $1-
|\braket{{\psi}_{1}^{in}}{{\psi}_{2}^{in}}|$

In order to compare the results obtained using the LIGe with the results
obtained when using the nonlinear transformation, one has to think about
how the LIGe can be used when two copies of the initial state are
available. It turns
out that there are two possibilities: $(1)$ perform two independent LIGe
measurements, one on each copy of the state or $(2)$ perform one single
measurement on the product state $\ket{{\psi}_{i}^{in}} \otimes
\ket{{\psi}_{i}^{in}}$. 

In the first case the probability of successfully identifying the state is
the probability of success in the first measurement plus the probability
of
failure in the first times success in the second, i.e. 
\begin{eqnarray}
{\rm Pr}_{\rm T}^{2 \times LIGe} (success) & = & {\rm Pr} (success)+
{\rm  Pr}(failure)\times {\rm Pr}(success)\nonumber\\
& = & 1 -
{|\braket{{\psi}_{1}^{in}}{{\psi}_{2}^{in}}|}^{2}
\end{eqnarray}

In the second case there one single LIGe measurement is performed on the
product state  $\ket{{\psi}_{i}^{in}} \otimes
\ket{{\psi}_{i}^{in}}$, the
probability
of successfully identifying the state is  
\begin{eqnarray}
{\rm Pr}_{T}^{1 \times LIGe}(success)= 1 - 
{|\braket{{\psi}_{1}^{in}}{{\psi}_{2}^{in}}|}^{2}
\end{eqnarray}
since the overlap between the two product states $\ket{{\psi}_{1}^{in}}
\otimes
\ket{{\psi}_{1}^{in}}$ and $\ket{{\psi}_{2}^{in}} \otimes
\ket{{\psi}_{2}^{in}}$ is equal  to
${|\braket{{\psi}_{1}^{in}}{{\psi}_{2}^{in}}|}^{2}$.
 
Notice that the three methods for state identification which have been
presented so far, all have the the same probability of success. In the
next section it is proven that this is indeed also the optimal solution.

\section{The optimal solution to the state identification problem}
\label{sec:opti}
In general when one wishes to distinguish deterministically between two 
non-orthogonal states and no errors are accepted, one is forced to 
introduce inconclusive answers. This means that there are three possible
outcomes, namely; the state was $\ket{{\psi}_{1}^{in}}$, the state  was
$\ket{{\psi}_{2}^{in}}$ or "don't know". A "don't know" means that that 
state was not successfully identified and the result is discarded in
order not to introduce any errors. This kind of measurement
is realized by what is called a Positive-Operator Value Measure (POVM)
\cite{peres}. 

The optimal POVM which answers these question is constructed in the
following way; the two projection operators ${\Bbb P}_{\neg
\ket{{\psi}_{1}^{in}}}={\openone}-
\ket{{\psi}_{1}^{in}}\bra{{\psi}_{1}^{in}}$ 
 ${\Bbb P}_{\neg
\ket{{\psi}_{2}^{in}}}={\openone}
-\ket{{\psi}_{2}^{in}}\bra{{\psi}_{2}^{in}}$
projects onto states
orthogonal to $\ket{{\psi}_{1}^{in}}$ and $\ket{{\psi}_{2}^{in}}$,
respectively. The three positive-operators which are needed (one of each
possible answer) are formed using these two projection operators,
\begin{eqnarray}
 & &{\Bbb A}_{\ket{{\psi}_{1}^{in}}}= x \left({\openone}
-\ket{{\psi}_{2}^{in}}\bra{{\psi}_{2}^{in}} \right) \nonumber \\
& &{\Bbb A}_{\ket{{\psi}_{2}^{in}}}= x \left({\openone}
 -\ket{{\psi}_{1}^{in}}\bra{{\psi}_{1}^{in}} \right) \\
 & &{\Bbb A}_{?}= {\openone} - {\Bbb A}_{\ket{{\psi}_{1}^{in}}}-{\Bbb
A}_{\ket{{\psi}_{2}^{in}}} \nonumber 
\end{eqnarray}
where the coefficient $x$ now is to be optimized. The first two
operators have the same coefficient because the initial states are
equi-probable. The requirement is now that the probability of an
inconclusive answer should be as low as possible, and that all three
operators must be positive. These requirements leads to the following
value,
\begin{eqnarray}
x=\frac{1}{1+{|\braket{{\psi}_{1}^{in}}{{\psi}_{2}^{in}}|}}
\end{eqnarray}
and gives probability ${|\braket{{\psi}_{1}^{in}}{{\psi}_{2}^{in}}|}$ of
obtaining an inconclusive answer. Hence the probability of successfully
determining the state is 
\begin{eqnarray}
{\rm Pr}(success)=1-{|\braket{{\psi}_{1}^{in}}{{\psi}_{2}^{in}}|}
\end{eqnarray}
 
Suppose now that two copies of the initial state is available, then the
probability
of successfully determining the state is 
\begin{eqnarray}
{\rm Pr}(success)=1-{|\braket{{\psi}_{1}^{in}}{{\psi}_{2}^{in}}|}^{2}
\end{eqnarray}
since the two copies can be thought of as the state
$\ket{{\psi}_{1}^{in}}\otimes \ket{{\psi}_{1}^{in}}$ or the state
$\ket{{\psi}_{2}^{in}}\otimes \ket{{\psi}_{2}^{in}}$, and these two state
have overlap ${|\braket{{\psi}_{1}^{in}}{{\psi}_{2}^{in}}|}^{2}$.

This shows that the methods for state identification which have been
presented in Sec.~\ref{sec:trid} and Sec.~\ref{sec:lige} are indeed
optimal.

\section{Purification of mixed states of spin-$1/2$}
\label{sec:puri}
It is straightforward to generalize the nonlinear transformations
described in Sec.~\ref{sec:how},
to apply them to mixed states of entangled pairs of spin-$1/2$. The
transformations can then be used to construct a purification scheme.
As previously, the idea is to have two physical systems in the same state
${\r}^{in}$, where ${\r}^{in}$ now represents an entangled pair of
spin-1/2, i.e. it is a $4\times 4$ density matrix. For
concreteness it is assumed that within each pair, one spin is carried by a
particle flying towards the left, while the other one is carried by 
a particle flying towards the right. The generalization consists in 
performing independently similar
operations as in (\ref{eq:trans}) to the two spins on the left hand side
(known as Alice) 
and to the two on the right hand side (known as Bob). The operation is
nearly identical to the XOR defined in (\ref{eq:xor}), with a sign
change on Bob's side:
\begin{eqnarray}
{{\Bbb {U}}_{A  }}=
\left( \begin{array}{cc} -i \spiny & 0\\ 0 & \openone \end{array} \right)
 ~~~
{{\Bbb {U}}_{B}}=
\left( \begin{array}{cc} i\spiny & 0\\ 0 & \openone \end{array} \right)  
\end{eqnarray}
The filtering is done, as in the single spin case, by selecting the spin
"down"
state of each of the spins in the target pair. This leads to the following
transformation:
\begin{eqnarray}
& & {\left({\openone}\otimes{\Bbb P}_{-}\right)}_{A} 
{\left({\openone}\otimes{\Bbb P}_{-}\right)}_{B} 
{\Bbb U}_{A} 
{\Bbb U}_{B}
{\left({\r}^{in}\otimes{\r}^{in}\right)}
{\Bbb U}_{B}^{\dag}
{\Bbb U}_{A}^{\dag} 
{\left({\openone}\otimes{\Bbb P}_{-}\right)}_{B}
{\left({\openone}\otimes{\Bbb P}_{-}\right)}_{A}\nonumber \\
& &={\r}^{out}\otimes{\Bbb P}_{--}
\end{eqnarray}
and gives  rise to the following outgoing density matrix shared between
Alice and Bob

\begin{eqnarray}
{\r}^{in}~~~\longrightarrow~~~{\rho}^{out} = \left( \begin{array}{rrrr} 
{({\rho}_{11})}^{2}&-{({\rho}_{12})}^{2}& 
{({\r}_{13})}^{2} &-{({\r}_{14})}^{2}\\
-{({\rho}_{21})}^{2} &{({\rho}_{22})}^{2}& 
-{({\r}_{23})}^{2}&{({\r}_{24})}^{2}\\
{({\r}_{31})}^{2}&-{({\r}_{32})}^{2}&
{({\r}_{33})}^{2}& -{({\r}_{34})}^{2}\\
-{({\r}_{41})}^{2}&{({\r}_{42})}^{2}&
-{({\r}_{43})}^{2}&{({\r}_{44})}^{2}
\end{array} \right)
\end{eqnarray}
This transformation preserves the the singlet state ${\psi}^{-}$. After
the transformation Alice and Bob both perform a bilateral (i.e. on both
sides) $\pi /2$
rotation
around the $x$-axis of their remaining spin. This rotation interchanges
the ${\psi}^{+}$ and the  ${\phi}^{+}$ Bell states without affecting the
other two Bell states. What they have hereby obtained is a quantum state
purification scheme which purifies towards the singlet state, ${\psi}^{-}$.

A purification scheme works in the following way: suppose the initial
state ${\r}^{in}$ had fidelity
\begin{eqnarray}
{F}^{in}={\bra{{\psi}^{-}}}{\r}^{in}{\ket{{\psi}^{-}}}
\end{eqnarray}
with respect to the singlet state. Taking two copies of ${\r}^{in}$,
performing the transformation and afterwards the rotation, the fidelity 
${F}^{out}_{rot}$ of the new state ${\r}^{out}_{rot}$ is bigger than
${F}^{in}$, i.e. ${F}^{out}_{rot}>{F}^{in}$. When the fidelity is equal
to 1 it means that the state is a pure state and therefore not entangled
with the environment (or an eavesdropper) \cite{bennett,deutsch}. 

Repeating the above operations, including the bilateral rotation, on
a state with fidelity ${F}^{in}>\frac{1}{2}$ selects a
subensamble with larger fidelity ${F}^{out}>{F}^{in}$. 
For example, suppose the initial state state had
fidelity ${F}^{in}=0.51$, after 10 iterations the fidelity is
${F}^{(10)}=0.809$ and after 15 iterations the fidelity is
${F}^{(15)}=0.99997$. It should be mentioned that, depending on the input
fidelity, ${F}^{in}$, the fidelity after the purification, ${F}^{out}$, 
may actually {\it decrease} for the first few iterations, but it will  
increase afterwards.

In order to have a higher efficiency (keep more pairs), Alice and Bob can
also keep the source
pair when the outcome of their measurement on the target pair gave them
$++$, in other words Alice and Bob can keep their source pair if they
both find $-$ or they both find $+$ when measuring their target pair. 

This purification scheme is, up to a phase, identical to the one developed
by Deutsch et al. \cite{deutsch}. The only difference being that in their
scheme the preserved state is the ${\phi}^{+}$ state, whereas here the
preserved state is the singlet state.

\section{Concluding remarks}
\label{sec:conre}
It has been shown that a non-linear quantum state transformation which
operates on pairs of spin-1/2, can be used to distinguish
deterministically
between two non-orthogonal states, provided two copies of the initial
states are available. The transformation, which involves only a unitary
operation (here the quantum XOR was used) and a filtering process (a
measurement), can actually transform
non-orthogonal states into orthogonal states. These states can now be
separated deterministically with a standard von
Neuman measurement. This transformation does not conflict with the 
basic laws for
quantum mechanics, which tells us that non-orthogonal states can not be
distinguished with certainty, since it only has a certain
probability of success, which then becomes the probability of successfully
determining the state.  

The result obtained when applying the non-linear transformation to the
state identification problem was compared with a specific POVM measurement
known as the LIGe (Loss Induced Generalized) quantum measurement. When two
copies of the initial state are provided the LIGe can be applied in two
different ways: either two independent LIGe measurements (one on each
copy) or a single measurement on the product state of the two copies. Both
procedures lead to the same probability for successfully determining the
initial state. The same probability was
obtained when using the non-linear transformation. This is in fact the
optimal solution to the state
identification problem, when inconclusive results --- but no errors ---
are accepted.

Finally it was shown how similar
transformations applied locally on each component of an entangled pair of
spin-1/2 can be used
to transform a mixed nonlocal state into a quasi-pure maximally entangled
singlet state. 

It should be mentioned that it is not only possible to square each
component of the density matrix as was seen in eq. (\ref{eq:sqrho}), but
it
can be raised to any power $n+1$. This is done by taking $n+1$ copies of
${\r}^{in}$, where $n$ of the copies act as target spins. A generalized XOR
is then applied, which flips the target spins if and only if the source
spin is spin-up. This is followed by a projection onto the spin-down of
all the target spins. This operation results in a density matrix
${\r}^{out}$ of the
source spin where each component has been raised to the power $n+1$.

It is also possible to extent the 'squaring' of the components of the
density matrix to higher dimensions. Suppose the initial pure state has
dimension $n$, i.e. $\ket{\psi}=({\psi}_{1},...,{\psi}_{n})$. Making
the tensor product of two identical states gives the new state,
$\ket{\psi}\otimes \ket{\psi}=({\psi}_{11},{\psi}_{12},...,{\psi}_{nn})$
with the
elements 
${\psi}_{11}={({\psi}_{1})}^{2}$, ${\psi}_{12}={\psi}_{1}{\psi}_{2}$, ...,  
${\psi}_{nn}={({\psi}_{n})}^{2}$. In order to select the squared elements,
i.e. ${\psi}_{ii}={({\psi}_{i})}^{2}$ the product state is rotated so that 
${\psi}_{11}\longrightarrow {\psi}_{1n}$, 
${\psi}_{22}\longrightarrow{\psi}_{2n}$, 
... , ${\psi}_{nn}\longrightarrow{\psi}_{nn}$. These are the elements
of
interest. The other elements can be rotated in an arbitrary way, as long
as they are not transformed into ${\psi}_{jn}$ for all $j$. A projection
onto the
spin-$n$ component of the target state leaves the target spin in
the spin-n state, where as the source spin is left in the state
where each  component has  been squared by itself. 

\section*{Acknowledgement}
We would like to thank Asher Peres for useful comments.
H.B.-P. is supported by the Danish National Science Research Council
(grant no. 9601645), B.H. is supported by the TMR network on Physics of
Quantum Information. This work is supported by the Swiss Fonds National de
Recherche Scientifique.

\newpage
\pagestyle{empty}
\newpage
\pagestyle{empty}
\samepage{
\begin{figure}[t]
\begin{center}   
\begin{tabular}{ccc}
&
\leavevmode
\hbox{%
\epsfxsize=3.3in
\epsffile{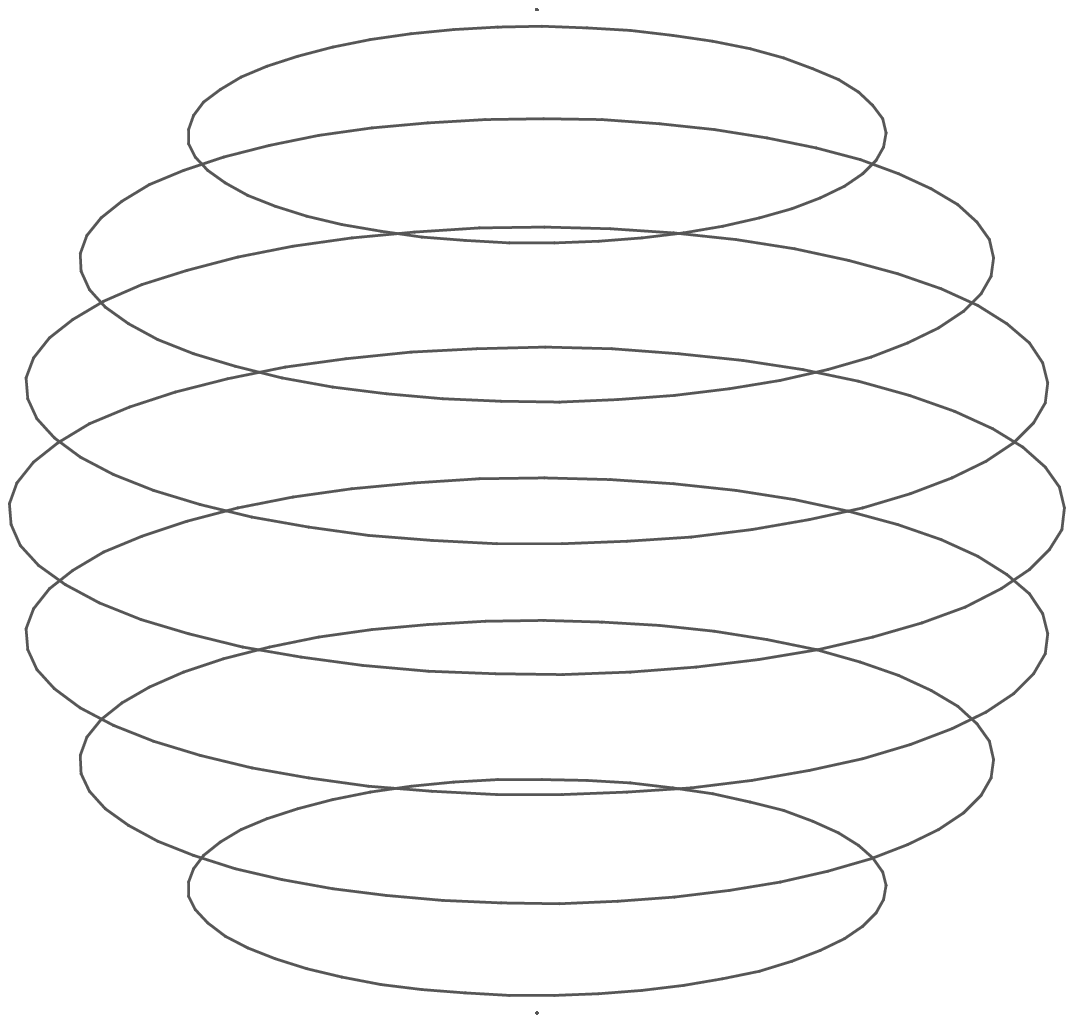}}& \\
& 
\leavevmode  
\hbox{%
\epsfxsize=2.2in
\epsffile{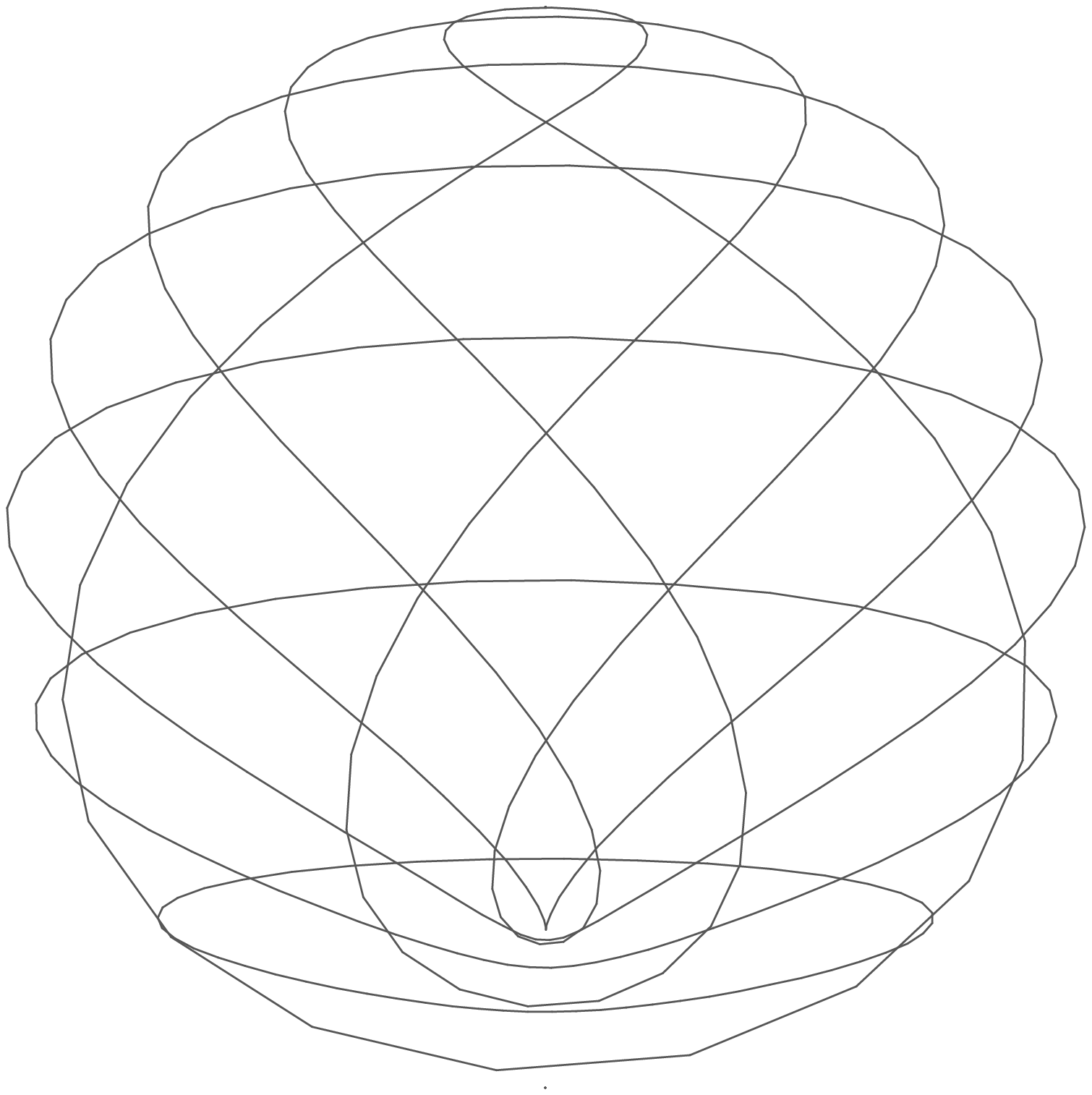}}&
\end{tabular}
\caption{The transformation of the sphere.
The upper figure shows the
spin-1/2 states represented on the
sphere in terms of their polarization or Bloch vector
(see Sec.~\ref{sec:trid}). A fully painted
sphere corresponds to all spin states. The lower figure shows the
transformed spin states. Here the
unitary operator used is not the XOR, but the operator ${\Bbb
U}={\exp({i\frac{\pi}{8}}{\sigma}_{z}\otimes {\sigma}_{x})}$.
} 
\end{center}  
\label{fig:sph1}
\end{figure}}


\begin{thebibliography}{99}
\bibitem{gisin1}N. Gisin, Phys. Lett. A {\bf 143}, (1990) 1
\bibitem{gisin2}N. Gisin, Helv. Phys. Acta, {\bf 62} (1989) 363
\bibitem{barenco}A. Barenco, D. Deutsch, A. Ekert and R. Jozsa, Phys. Rev.
Lett. {\bf 74}, (1995) 4083
\bibitem{ivan}I. D. Ivanovic, Phys. Lett. A {\bf 123}, (1987) 257 
\bibitem{peres1}A. Peres, Phys. Lett. A {\bf 128}, (1988) 19 
\bibitem{huttner}B. Huttner, A. Muller, J. D. Gautier, H. Zbinden and N. 
Gisin, Phys. Rev. A, {\bf 54}, (1996) 3783
\bibitem{peres}A. Peres, "Quantum Theory: Concepts and Methods" (Kluwer,
Dordrecht, 1993)
\bibitem{bennett}C. H. Bennett, G. Brassard, S. Popescu, B. Schumacher,
J. Smolin and W. Wootters, Phys. Rev. Lett, {\bf 76}, (1996) 722
\bibitem{deutsch}D. Deutsch, A Ekert, R. Jozsa, C. Macchiavello, S.
Popescu and A. Sanpera, Phys. Rev. Lett, {\bf 77}, (1996) 2818


\end{thebibliography}
\end{document}